\documentclass[12pt,a4paper]{article}
\usepackage[latin1]{inputenc}
\usepackage[english]{babel}
\usepackage{verbatim}
\usepackage[dvips]{graphicx}
\usepackage{amstext}
\usepackage{amsfonts}
\usepackage{latexsym}
\usepackage{amsmath}
\usepackage{geometry}
\usepackage{amsmath}
\usepackage[english]{babel}
\usepackage[dvips]{graphicx}
\usepackage[latin1]{inputenc}
\usepackage{amsmath,amssymb,amsthm,upref}
\newcommand{\e}{ \,\text{e}[}
\newcommand{\alp}{|}
\newcommand{\re}{]}

\newtheorem{lemma}{Lemma}%[section*]
\newtheorem{definition}{Definition}%[section*]
\begin{document}
\author{Peter Nyman\\ \\International Center for Mathematical Modeling\\in Physics, Engineering and Cognitive science\\MSI, Växjö University, S-35195, Sweden} 
\title{Simulation of Quantum Algorithms with a Symbolic Programming Language}     
%\runtitle{Simulation of Quantum Algorithms with a Symbolic Programming Language}
%\runauthor{Peter Nyman}
%\classification{03.67.Lx, 02.70.-c, 07.05.Tp}
%\keywords      {Shor's factoring algorithm, prime factorization, quantum computing, \textit{Mathematica}, quantum Fourier transform}
\maketitle 
%#####################
%                     \input{Abstract}
%#####################
\begin{abstract}
This study examines the simulation of quantum algorithms on a classical computer. The program code implemented on a classical computer will be a straight connection between the mathematical formulation of quantum mechanics and computational methods. The computational language will include formulations such as quantum state, superposition and quantum operator.
\end{abstract}
%#####################
%                     \input{Introduction}
%#####################
\section{Introduction}
The history of quantum computing begins around the 1980s when Richard Feynman at the First Conference on the Physics of Computation showed that it is not effective to simulate an evolution of a quantum system on a classical computer. An effective simulation of quantum system has a run-time in polynomial size, i.e. the computational time is smaller than a polynomial function of the problem size. The conclusion is that relevant simulations of quantum computers will always be larger in size than polynomial time. This will bring our attention to the super-polynomial time simulations of quantum algorithms; these kinds of simulations have a long run-time for large problem. By separating the problems in smaller parts we can avoid the long run-time. For examples, there will be super-polynomial time to simulate \textit{Shor's factoring algorithm} on a classical computer. The simulation of quantum algorithms will still be constructive for parts of a larger problem and it will give us a basis for comparison between experimental results and theoretical results. 
The results from \textit{Shor's  algorithm} will be verified by multiple the factors from the algorithm outcome and hence it is straightforward to check the results from \textit{Shor's factoring algorithm} implemented on a quantum computer.
It might be more complicated to check the outputs from future algorithms.
However, it is possible to show that \textit{Shor's algorithm} will give mathematically correct results, see \cite{Shor:2001}.
But how can we verify that implementations of \textit{Shor's algorithm} on a quantum computer which coincides with their mathematical model? A simulation of a quantum algorithm in classical computer will give us the possibility to compare the outcome from a quantum computer with the output form a physically more stable classical computer. In the development of quantum algorithms it will be interesting to check new algorithms on a classical computer. This article will show a connection between a future quantum computer and today's simulations of quantum computers. The intention is to demonstrate a computational simulation of quantum algorithms on a classical computer. A couple of examples of various methods for the simulation of quantum algorithms were given in \cite{Julia:2006,Altenkirch:2004}. One of these is the Mathematica package Qdensity which is a simulation of a quantum computer developed by B.Juliá-Díaz,  J.M. Burdis and F. Tabakin. Qdensity gives a simulation of a number of well-known quantum algorithms such as teleportation, Shor's algorithm, Grover's search and more. 
An operation in Qdensity is matrix operation and therefore matrixes will represent quantum gates and column matrixes will represent a qubit.  The package includes the most up-to-date essential operations in quantum computing. As examples of these operations we will mention the \textit{Hadamard}, \textit{controlled not} and \textit{Toffoli} gates.
 In preprint \cite{Nyman:2005} we introduced a computational language constructed on the basis of quantum mechanics. We have decided to implement the well-known \textit{Shor's algorithm} as an example of a quantum algorithm. The aim is to construct a computational language in order to present a straightforward connection between \textit{Dirac's} mathematical formulation of quantum mechanics and the program code. Thus, the computational language will include quantum mechanical terminology such as quantum operators and quantum states. A mathematically described algorithm will have a computational algorithm that has a clear mathematical structure in the program code. This simulation will be a link from the quantum algorithm described in terms of mathematical concept to the implementation of this quantum algorithm on a quantum computer. First we will construct a simulation framework in the high-level program language \textit{Mathematica}.  
 
 %#####################
%                     \input{BasisState}
%#####################
\section{The Simulation Framework}\label{Basicstate}
This section will introduce a framework construct for the simulation of quantum algorithms in classical computers. We wish to point out that there is a symbolic similarity between our framework and the mathematical framework. This framework will be a computational dual to \textit{Dirac's bra-ket notation}. 
A quantum state in $n$ dimensions can be represented by a linear
combination of $n$ numbers of basis vectors. In the two-dimensional
case a quantum state $|\phi\rangle$ is represented as a
superposition of two basis vectors, say $|0\rangle$ and $|1\rangle,$
(computation basis \cite{Mika:2001,Isaac:2000}). In this case a
quantum state $|\phi\rangle$ is represented as
\begin{equation}\label{phi}|\phi\rangle=\alpha|0\rangle+\beta|1\rangle,
\end{equation}
where $\alpha$ and $\beta$ are complex numbers that have the sum of
squares equal to one.
We will introduce some new symbols for the states of this computational basis as follows:
$\text{e}[0]= |0\rangle$ and $\text{e}[1]= |1\rangle.$ This is the
foundation for the structure of the program code. For more than
one-qubit we will use the computational basis states
$\text{e}[x_1,\ldots,x_n]= |x_1\ldots x_n\rangle$, where
$x_j\in\{0,1\}$ or by  using the more compact notation $\text{e}[y]=
|y\rangle$, where $y=x_n 2^0+\dots+ x_1 2^{n-1} $. We will, write
the state $\phi$ as $ \text{e}[\phi]=\alpha \text{e}[0]+\beta
\text{e}[1]$, by analogy with the equation (\ref{phi}). The operator $A$
acts on the state $\phi$ and is often written as $A|\phi\rangle$ in
the quantum mechanical literature. To match these symbols, we will
use the computational symbols $A\alp \text{e}[\phi]$ for this
operation. A computational problem is to handle different expression with identical meaning. As an example the expression  $\text{e}[0,\text{e}[1],1]$ must be equal to $\text{e}[0,1,1]$ in the code. We can bring in the command $\text{e}[0,\text{e}[1],1]:=\text{e}[0,1,1]$ or the more general $ \text{e}[a\_\_,\text{e}[b\_\_],c\_\_]:=\text{e}[a,b,c]$  to solve this problem. Moreover the program code must be able to handle the linearity of
tensor product. Let $\text{e}[\,.\,]$ be vectors and
$\alpha$ a scalar. We define the tensor product as
\begin{eqnarray}\label{tensor1}
\alpha(\text{e}[v]\otimes \text{e}[w])=(\alpha\text{e}[v])\otimes\text{e}[w]=\text{e}[v]\otimes(\alpha\text{e}[w])\\
(\text{e}[v_1]+ \text{e}[v_2])\otimes\text{e}[w]=\text{e}[v_1]\otimes\text{e}[w]+ \text{e}[v_2]\otimes\text{e}[w]\\
\text{e}[v]\otimes(\text{e}[w_1]+ \text{e}[w_2])=\text{e}[v]\otimes\text{e}[w_1]+ \text{e}[v]\otimes\text{e}[w_2].
\end{eqnarray}
We add two commands to the program code that will implement this
definition of the tensor product. The command
\[\text{e}[a\_\_\,,\alpha\_\,. \text{e}[x\_\_\,],b\_\_\,]:=\alpha
\text{e}[a,x,b]\] will transform
$\text{e}[a]\otimes\alpha\text{e}[x]\otimes\text{e}[c]$  to
$\alpha\text{e}[a\otimes x\otimes b]=\alpha\text{e}[a,x,b]$. This command is the computational dual to the tensor expression in Dirac's notation  $ |a\rangle\otimes\alpha |x\rangle\otimes|b\rangle= \alpha |a\, x\, b\rangle$. The other command
\[\text{e}[a,\xi(\alpha\e x\re+\beta\e
y\re),b]:=\xi\alpha\text{e}[a,x,b]+\xi\beta\text{e}[a,y,b]\] will
transform $\text{e}[a]\otimes\xi(\alpha\e x\re+\beta\e
y\re)\otimes\text{e}[b]$ to $\xi\alpha\text{e}[a,x,b]+\xi\beta\text{e}[a,y,b]$. Let $U$ be an arbitrary one-qubit quantum
gate. Then $U$ will transform an arbitrary state $\e\phi\re$ which
is represented in the computational basis states as $\e\phi\re=a\e
0\re+ b\e 1\re $ to the state $U\alp   \e\phi\re\rightarrow\ a(c_1\e
0\re+ c_2\e 1\re )+ b(c_3\e 0\re+ c_4\e 1\re )$, where $a,b,c_i\in
\mathbb{C}$. Now we add a \textit{Mathematica} gate $U$ to the
program code, with the following result $U|\e0\re\rightarrow c_1\e 0\re+ c_2\e
1\re$ and   $U|\e1\re\rightarrow c_3\e 0\re+ c_4\e 1\re $. For
example, the Hadamard gate $H$ will be added in \textit{Mathematica}
as the command $H{:=}\{\e0\re\rightarrow 1/\sqrt{2}(\e 0\re+ \e
1\re),\e1\re\rightarrow 1/\sqrt{2}(\e 0\re- \e 1\re)\}$. We will
define a one-qubit gate $O_i$ as an operator which acts on the qubit
in position $i$ and leaves the other qubits unchanged. The program
code must be able to operate with a gate on an arbitrary qubit.
Consequently we will define an operator $O_i$ in the
\textit{Mathematica} code. The operator will be defined
$O_i=I^{\otimes i-1}\otimes U\otimes I^{\otimes n-i}$
%$O_i=\bigotimes_1^{i-1}I\otimes O\otimes\bigotimes_{i+1}^nI$
as an operator which acts on $n$-qubits,where
$I$ is the one-qubit unit operator and $U$ is an arbitrary one-qubit operator.
Then operator $O_i$ is a function of $O_i\alp \e v\re\rightarrow \e\psi\re$.
%Similarly we will defined $O_{i,j}\equiv I^{\otimes i-1}\otimes U_1\otimes I^{\otimes j-i-2}\otimes U_2\otimes I^{\otimes j-i-2}$, where $U=U_1\otimes U_2$ is a two-qubits operator.
Similarly, we will define $O_{i,j}$ as an operator  which operates
as the two-qubits operator on the qubits in positions $i,j$  and
leaves the other qubits unchanged. Now we have the tools to build
quantum circuits. To achieve this, we will use a quantum Fourier
transform circuit in Shor's factoring algorithm.
%#####################
%                     \input{IntroductionToShorsFactoringAlgorithm}
%#####################
\section{An Introduction to Shor's factoring algorithm}
Prime factorizing of an odd number $N$ can be accomplished  using
Shor's algorithm \cite{Shor:2001}. If $N$ is an even integer then we
divide it with $2$ $n$-times until $2^{-n}N$ becomes an odd number.
An even $N=2 n$ can easily be found in view of the fact that $2 n
\equiv0\pmod 2$.
%\begin{theorem}
Let N be the composite of prime factors so that
 \begin{equation}\label{primefactor1}
    N=p_1^{\alpha_1} p_2^{\alpha_2}\ldots p_k^{\alpha_k},
    \end{equation}
where  $k>1$ and $\alpha_i\in\mathbb{Z}^+$.
%\end{theorem}
The algorithm will be able to factorize the integer $N$. We can also
assume that $N$ is not a  prime power, i.e. $k>1$ and that there
exists at least one $p_i\neq p_j$. A prime power can be found with a
known classical method in polynomial time. Let us choose a $x\in
\mathbb{Z}_N$ randomly, where $\mathbb{Z}_N=\{1,2,\ldots,N-1,N\}$.
The next step is to use Euclid's algorithm which determines the
greatest common divisor. If  $x$ and $N$ are not relatively prime,
then we will find a factor by using Euclid's algorithm. A factor is
equal to $\text{gcd}(x,N)$ if $\text{gcd}(x,N)\neq1$.
  If  $x$ and $N$ are relatively prime, the task will be to find the
order of $x$ in the group $\mathbb{Z}_N$.
The algorithm will search for the smallest $r\in\mathbb{Z}^+$ so
that $x^r\equiv1 \pmod N$; consequently $r$ is called order $r$ of $x$. %Then the notation$r=ord_N(x)$ means the order of element $x$ in $\mathbb{Z}_N$. %See chapter \ref{order} for the definition of order.
If
 \begin{equation}\label{primefactor}
x^r\equiv1 \pmod N
  \end{equation}
and $r$ is a even integer, then it is possible to factorize
$x^{r}-1$ as
    \begin{equation}\label{factorupp}
    x^r-1=(x^\frac{r}{2}-1)(x^\frac{r}{2}+1).
    \end{equation}
%It is possible to prov that it exist a unique solution for $x$ in equation (\ref{primefactor}) by using the Chinese remainder theorem.
The integer $N$ will share at least one factor with
$(x^\frac{r}{2}-1)$ or $(x^\frac{r}{2}+1)$ since $N$ divides $x^r-1$. These factors can be calculated with Euclid's algorithm. Let us presume that an even $r$ has been found, which gives us the possibility to determine the factors equal to
$\gcd(x^\frac{r}{2}+1,N) $ and $\gcd (x^\frac{r}{2} -1 , N)$.
There is nevertheless an obvious exclusion
that $x^\frac{r}{2}\equiv1  \pmod N $ due to the definition of order $r$. However, it may occur that $x^\frac{r}{2}\equiv-1  \pmod N $, i.e. only trivial factors will be found since $N$ is the greatest common divisor to $x^\frac{r}{2}+1$ and $N$. There will be two factors, $\gcd(x^\frac{r}{2}+1,N) $ and $\gcd (x^\frac{r}{2} -1 , N)$, in case the order $r$ of $x$ is an even number and $x^\frac{r}{2}\not\equiv -1 \pmod N $.
%#####################
%                     \input{QFT}
%#####################
\section{The Quantum Fourier Transform}\label{Quantum Fourier Transform}
%\subsection{Quantum Fourier Transform}
The Quantum Fourier transform is a discrete Fourier transform procedure applied to the framework of quantum mechanics, used to realize \textit{Shor's  algorithm} and some other really interesting quantum algorithms. 
The discrete Fourier transform maps a vector $V_1=\alpha_0, \alpha_1,\ldots,
\alpha_{n-1}$ to another vector  $V_2=\beta_0, \beta _1,\ldots\beta _k\ldots, \beta
_{n-1}$, where $\alpha,\beta\in\mathbb{C}$. In traditional mathematics the discrete
Fourier transform is defined as follow:
\begin{equation}\label{V2}
V_2=\frac{1}{\sqrt{n}}\sum_{j=0}^{n-1}\alpha_j e^{2\pi i j k/n}.
\end{equation}
 The quantum Fourier transform (QFT) is analogue to the discrete Fourier transform, yet QFT will be represented in
computational symbols. The QFT is a function which acts on $q=\log_2 n$
qubits.
\begin{definition}\label{DEFQFT} The quantum Fourier transform is a function which maps basis states to a linear combination of basis states
{\rm
\begin{equation}\label{QFT}QFT|\e j\re=\frac{1}{\sqrt{n}}\sum_{k=0}^{n-1} e^{2\pi i j k/n}\e k\re .\end{equation}
} The quantum Fourier transform for an arbitrary state $\psi$ is
{\rm
\begin{equation}\label{arbitrary linear combinations}
 QFT|\sum_{j=0}^{n-1} \alpha_j\e j\re=\frac{1}{\sqrt{n}}\sum_{k=0}^{n-1}\alpha_k e^{2\pi i j
 k/n}\e k\re.
\end{equation}
}
\end{definition} 
A decomposition of QFT will be necessary in the task in order to implement the circuit on a quantum computer; therefore we will work with the decomposition of QFT in our simulation.  This decomposition is one of the essential items in the creation of a QFT circuit, since it will allow us to construct the circuit with the use of quantum gates.
Let us introduce the decomposition  as
\begin{lemma}\label{lemmadecomposion}
{\rm
\begin{equation}\label{decomposition}
 QFT|\e j\re=\frac{1}{\sqrt{n}}(\e 0\re+e^{\frac{2\pi i
j}{2^1}}\e 1\re)\otimes(\e 0\re+e^{\frac{2\pi i
j}{2^2}}\e 1\re)\otimes\cdots\otimes(\e 0\re+e^{\frac{2\pi i j}{n}}\e 1\re).
\end{equation}
}
\end{lemma}
%=================================================================================
%============================\begin{lemma}\label{lemmadecomposion}
%=================================================================================

\begin{figure}
\centering
\includegraphics{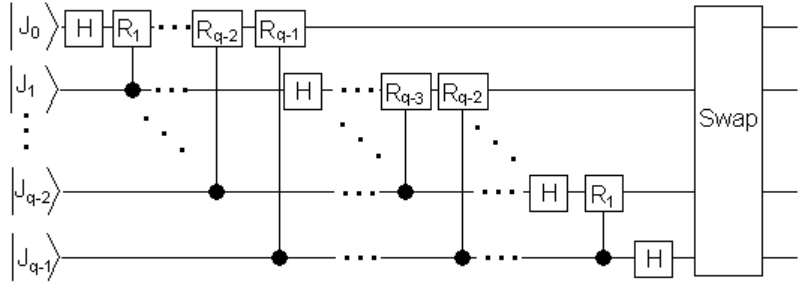}
\caption{Circuit for the Quantum Fourier Transform}\label{QFTfigure}
\end{figure}

We will use the combination of the compact notation $\text{e}[y]$ and the notation $\e x_1,\ldots ,x_n\re $ where $y=%\sum_{i=1}^n x_i 2^{n-i}=
x_n 2^0+\dots+ x_1 2^{n-1} $ and
$x_j\in\{0,1\}$. Hence, we write $e[y,1]$ and $e[y,0]$ with the meaning $\e x_1,\ldots ,x_{n-1},1\re $ and $\e x_1,\ldots ,x_{n-1},0\re $ respectively.
The proof of this lemma will follow from the fact that  $ \e l, 0\re \in \{\e  0\re, \e  2\re,\e  4\re ,\ldots\}$  and $ \e l,1\re \in \{\e  1\re, \e  3\re,\e  5\re ,\ldots\}$. 
\proof
 \begin{eqnarray}QFT|\e j\re&=&\frac{1}{\sqrt{n}}\sum_{k=0}^{n-1} e^{2\pi i j
 k/n}\e k\re\\
&=&\frac{1}{\sqrt{n}}(\sum_{l=0}^{n/2-1} e^{2\pi i j
2l/n}\e l, 0\re+\sum_{l=0}^{n/2-1} e^{2\pi i j (2l+1)/n}\e l,1\re)\nonumber\\
&=&\frac{1}{\sqrt{n}}(\sum_{l=0}^{n/2-1}e^{2\pi i j2l/n}\e l, 0\re+\sum_{l=0}^{n/2-1}
e^{2\pi i j 2l/n} e^{2\pi ij/n}\e l, 1\re)\nonumber\\
&=&\frac{1}{\sqrt{n}}((\sum_{k=0}^{n/2-1}e^{2\pi i j
k/(n2^{-1})}\e k\re)\otimes(\e 0\re+e^{\frac{2\pi i j}{n}}\e 1\re))\nonumber\\
&=&\frac{1}{\sqrt{n}}((\sum_{k=0}^{n/4-1}e^{2\pi i j
k/(n2^{-2})}\e k\re)\otimes(\e 0\re+e^{\frac{2\pi i j}{n2^{-1}}}\e 1\re)\otimes(\e 0\re+e^{\frac{2\pi i j}{n}}\e 1\re))\nonumber
  \end{eqnarray}
The proof will follow if we continue with this method and decompose the remaining qubits in a similarly way.
However, the Rotation, the Hadamard and the CNOT gates will realize the decomposition in the circuit simulation, illustrated in Figure \ref{QFTfigure}.
%====================circuit1
%====================circuit1 
 Every arbitrary unitary operator may be represented by combinations of single qubit gates and CNOT gates, see \cite {Isaac:2000}. %SID189
QFT can be expressed in single quantum gates as:
\begin{equation}\label{q-qubit}
 Swap(H_{q}(R_{q,q-1}H_{q-1}(\cdots(R_{q,2}\cdots
R_{3,2}H_{2}(R_{q,1}\cdots R_{2,1}H_{1}|\e\mathbf{u}\re))))),
\end{equation}
where swap is a combination of $3q$ numbers of CNOT gates.
%In equation \eqref{q-qubit} $H_{d}$ means that $H$ operate on the qubit
%in position $d$. Similar to this we mean that $R_k=R_{|a-b|}$ operate on the qubit in position
%$a$ and $b$ when we write $R_{a,b}$.
This decomposition of QFT requires $q$
operations on the first qubit and $q-1$ operations on the second qubit, and so on. Hence it follows that the decomposition needs  $\frac{1}{2}(q+1)q$ $H$ and $R_k$ operations. To obtain the right order, we swap the decomposition; thus we need $3q/2$ or $3(q-1)/2$ more
operations. Altogether, the decomposition of QFT is requires in the order of $q^2$ gates e.i.
QFT uses $O(q^2)$ elementary operations.
%#####################
%                     \input{QuantumComputationAndShorsFactoringAlgorithm}
%#####################
\section{Quantum computation and Shor's factoring algorithm.}\label{shorsalgo}
Shor's algorithm will be divided in four steps, the first, second and last step can be executed on a classical computer in polynomial time and we will therefore be including these steps as classical algorithm in the simulation.  First let us choose an
arbitrary integer $x\in \mathbb{Z}^+$ which will be smaller than the integer $N$ that we want to factorize.
The second step is to make sure that the chosen integer is not a prime factor; otherwise the algorithm has found a factor. It is only in the third step we need to implement the algorithms on a quantum computer. In this stage we will use the quantum Fourier transform in the algorithm to find the order $r$ of $x$. Restart the algorithm if $r$ is odd or $x^\frac{r}{2}\equiv -1 \pmod N$. In the last stage a classical computer will calculate the factors and output $\gcd( {x^\frac{r}{2}\pm1, N}) $.
Let us start the algorithm with an $N$  and choose $n=2^q$ so that $N^2 \leq n \leq
2N^2$. It will prepare two registers $\e\textbf{0}\re\e\textbf{0}\re$ with $q$-
qubits in the quantum computer.
\begin{itemize}
  \item [1.]Let us set up the first register in superposition
        \begin{equation}\label{step1}
            \frac{1}{\sqrt{n}}\sum_{c=0}^{n-1}\e{c}\re\e\textbf{0}\re.
        \end{equation}
  \item [2.]Let us then compute $x^c \pmod N$ in the second register and the
  computer will be in state
          \begin{equation}\label{step2}
            \frac{1}{\sqrt{n}}\sum_{c=0}^{n-1}\e{c}\re\e{x}^{c} \text{ mod} N\re.
        \end{equation}
  \item [3.]Next, we need to compute the QFT on the first register $\e{c}\re$
  \begin{equation}
  \label{QF}QFT|\e{c}\re=\frac{1}{\sqrt{n}}\sum_{k=0}^{n-1} e^{2\pi i c k/n}\e{k}\re .
  \end{equation}
  The machine will be in state
   \begin{equation}\label{step3}
           \psi= \frac{1}{n}\sum_{c=0}^{n-1}\sum_{k=0}^{n-1}e^{2\pi i c k/n}\e{k}\re\e{x}^{c} \text{ mod} N\re.
    \end{equation}
\end{itemize}
%==================================================================
%(((((((((((((((((((((((((((((((((((((((((((((((((((((((((((((
%\\\\\\\\\\\\\\\\\\\\\\\\\\\\\\\\\\\\\\\\\\\\\\\\\\\\\\\\\\\\\\\\\\\
The algorithm will measure the first register for any $k$, but we calculate a measure of both registers, since it is the same probability to measure $k$ from the first register as there is to measure  $ \e{k},{x}^{c} \text{ mod} N \re$ from both of the registers.
The algorithm will find a prime factor only if $r$ is an even integer and the following condition is satisfied $x^\frac{r}{2}\neq -1 \pmod N $. The probability that a randomly chosen $x$ will satisfy these two conditions is at least $9/16$. If we choose  $n\geq N^2$ the quantum algorithm will find a useful $r$ with the probability that is at least $\frac{9}{160 \log \log N}$. The algorithm will be successful, with high probability if we execute this algorithm for $\frac{160 \log \log N}{9}$ times. 
%==================================================================
%(((((((((((((((((((((((((((((((((((((((((((((((((((((((((((((
%\\\\\\\\\\\\\\\\\\\\\\\\\\\\\\\\\\\\\\\\\\\\\\\\\\\\\\\\\\\\\\\\\\\
Now let us measure the first register in the quantum machine for any $k$ in $\e{k}\re$.
The order $r$ of $x$ can be found as a denominator of one of the convergents to $k/n$, where the probability to find $r$ depends on the number of qubits.
To find the order $r$, we need to apply continued fractions to the $k/n$
% Let $a$ be an integer so that $0 \leq a <r$, where $r$ is the order. %We can
%matematica code Antalupprepningar=\!\(\((9. \/\(160\ Log[Log[10. ^100]]\))\)\^\(-1\)\) och P=\!\(1 - \((1 - 9. \/\(160\ Log[Log[10. ^100]]\))\)\^97\)
%#####################
%                     \input{ShorsAlgorithmInMathematica}
%#####################
\section{Shor's Algorithm in the \textit{Mathematica} code}

This chapter will introduce a \textit{Mathematica} code which implements Shor's algorithm on a classical computer. We will follow the \textit{Mathematica} code evolutions and compare this with Shor's algorithm.
   This comparison will demonstrate a connection between the classical computer and the quantum computer. 
The program code is divided in parts with the comments below the code to make it more readable.
%We
%have drawn a box around \textit{Mathematica} program code, which makes it
%easier to distinguish the \textit{Mathematica} code from the comments.
The program will try to find two factors to $N$, where $N$ is an odd
composite of prime factors and has at least two different prime factors.
%=============================================================bild1
\begin{center}
\includegraphics{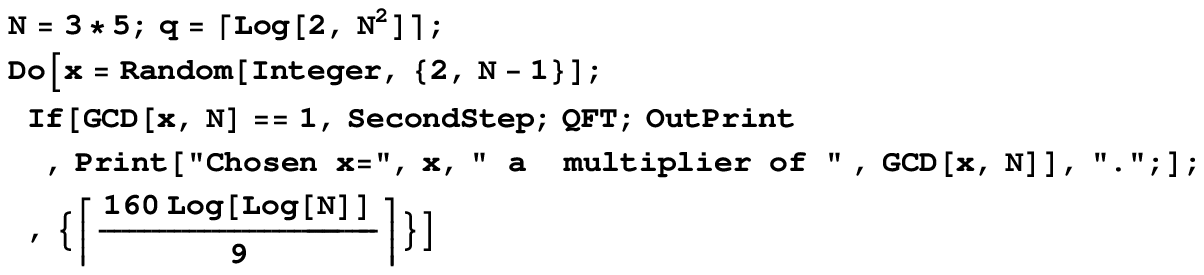}
\end{center}
The prospect for the algorithm to success will increase with numbers of qubits, but the execute time will also increase. Of course, this simulation will be a super- polynomial time algorithm and large integer will not be able to factorize in reasonable time.      
The algorithm will choose \textbf{q}$=\left\lceil  Log_2 (N^2)\right\rceil$ so that the algorithm will find a factor with large probability, i.e. if it selects $\textbf{q}$ to satisfy $N^2 \leq 2^\textbf{q}<2N^2$, the two factors will then be found with a probability of at least $\frac{9}{160 \log \log N}$. \\ The program will choose a random integer $\textbf{x}\in\{2,3,\cdots,N-1\}$. 
%=============================================================bild1
\begin{center}
\includegraphics{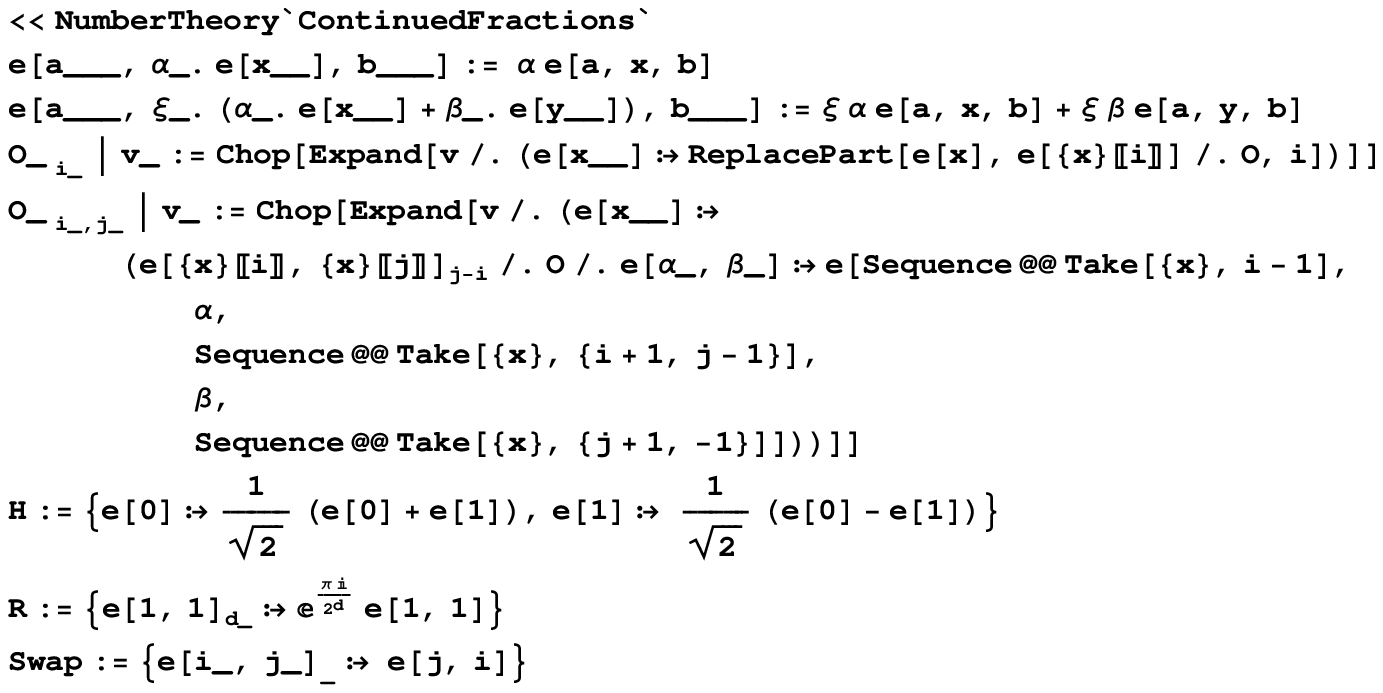}
\end{center}
We will only use the computational basis states
$\text{e}[x_1,\ldots,x_n]$, where $x_j\in\{0,1\}$.
% The code $\text{e}[a\_\_\,,\alpha\_\,.
%\text{e}[x\_\_\,],b\_\_\,]:=\alpha \text{e}[a,x,b]$ will give us
%the opportunity to perform tensor product to the two states say
%$a\text{e}[0]$ and $b\text{e}[1]$ so that  $a \text{e} [0] \otimes
%b \text{e}[1 ]=a[0,b\text{e}[1] ]= ab\text{e}[0,1 ]$. The code in
%next line take $\text{e}[a,\xi(\alpha\e x\re +\beta\e y\re),b]$ to
%$\xi\alpha\text{e}[a,x,b]+\xi\beta\text{e}[a,y,b]$ in similarly
%way.
The commands $\text{e}[a\_\_\,,\alpha\_\,. \text{e}[x\_\_\,],b\_\_\,]:=\alpha \text{e}[a,x,b]$ and $\text{e}[a,\xi(\alpha\e x\re+\beta\e y\re),b]:=\xi\alpha\text{e}[a,x,b]+\xi\beta\text{e}[a,y,b]$ will give the program code a connection to the tensor product.
The command $\textbf{O}\!\_\;_{i\_}|v\_$ is a one-qubit
operator which takes vector $v$ as an attribute and operates with
$O$ on the qubit in position $i$. Likewise, the command
$\textbf{O}\!\_\;_{i\_\;,j\_}|v\_$ is a two-qubits operator which
takes vector $v$ as an input and operates with $O$ on the qubit in
position $i$ and $j$. To compute QFT the algorithm requires three gates,
Hadamard (\textbf{H}), Rotation (\textbf{R}) and Swap
(\textbf{Swap}).
%The program will use the \textbf{Swap} gate, even if it is not necessarily. It is easier to read the qubit in opposite order instead of swap the qubit.
\begin{center}
%============================================================bild3
\includegraphics{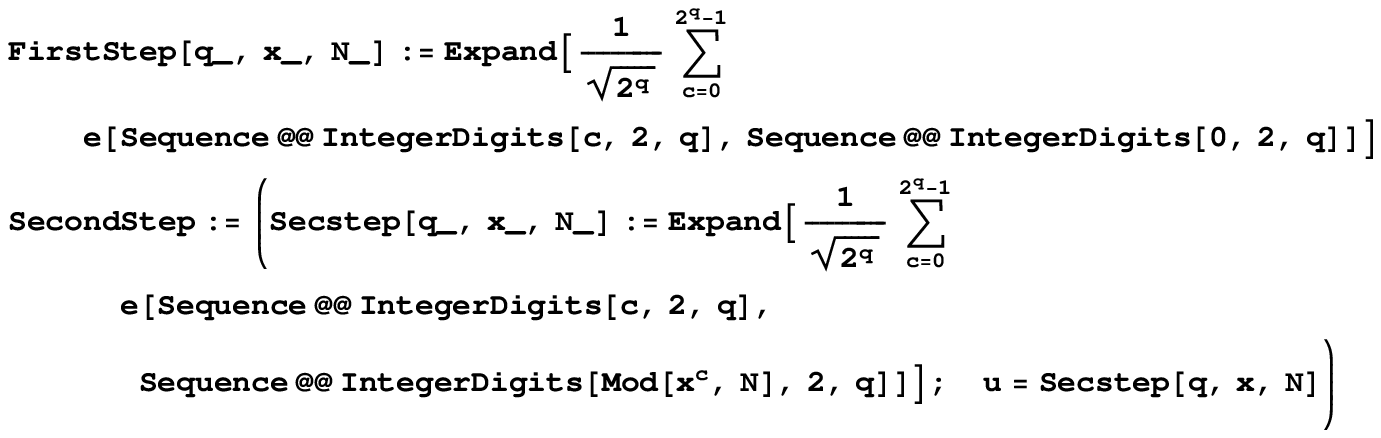}
\end{center}
The command \textbf{FirstStep} prepares the first register in a superposition. Since the first step is pointless on a classical computer; consequently it will be excluded in the code and we go directly to the second step in the code.
\textbf{Secstep} calculates $x^c \pmod N$ in the second register, where $q$ is the number of qubits.
%The number to factorize is $N$.
%One of all the number in the second register chose randomly by means of the function \textbf{RandomMeasure}, i.e. the function chose any of $x^b \pmod n$ there $b\in\{0,\cdots r-1\}$. \textbf{MeasureSecStep} select all term there $x^b \pmod n$ and make a new state. With use of the function \textbf{u} take this state to a normalize state in binary digits.
\begin{center}
%=============================================================bild4
\includegraphics{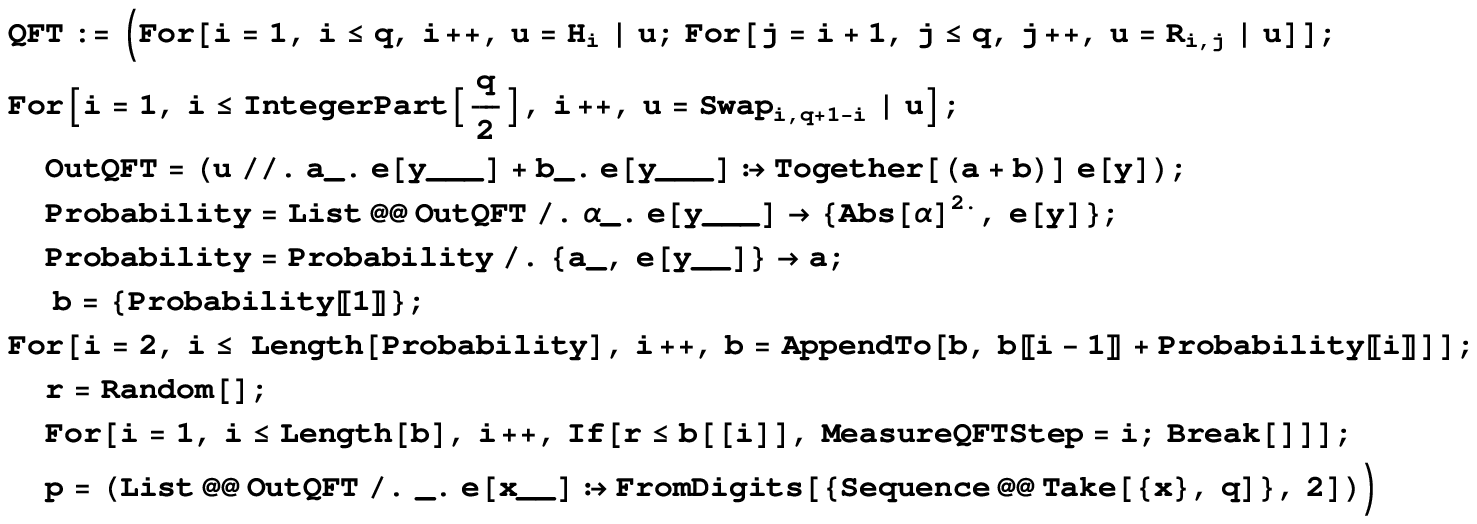}
\end{center}
The \textbf{QFT} will act on the state $u$ of the three gates in the following order: \\
$H_{q}(R_{q,q-1}H_{q-1}(\cdots(R_{q,2}\cdots
R_{3,2}H_{2}(R_{q,1}\cdots R_{2,1}H_{1}\e\mathbf{u}\re))))$. The third
line in the program code will swap the qubits. All terms with identical computational basis states will be collected in the command \textbf{OutQFT}.
\textbf{Probability} is a list of the probabilities used to measure one of the terms in the register.
One of the terms will be randomly chosen, taking into consideration of probability to measure the state. The position of the chosen term will be saved in \textbf{MeasureQFTStep}.
Finally, the list \textbf{p} of decimal numbers is derived from the binary list \textbf{OutQFT}.
%=============================================================bild6
\begin{center}
\includegraphics{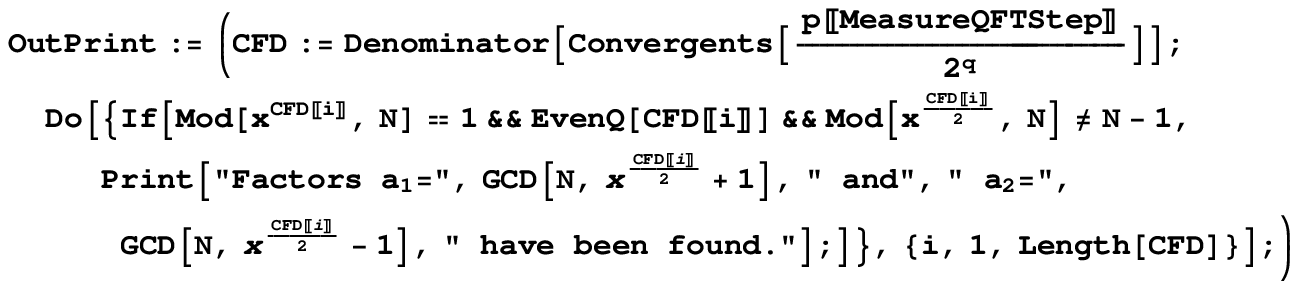}
\end{center}
The randomly chosen value in the register is in p[[\textbf{MeasureQFTStep}]].
In \textbf{CFD} the program saves the denominator of convergents ${\text{p[[\textbf{MeasureQFTStep}]]}}/{2^q}$. From this we can select all even denominators, where $\text{\textbf{x}}^\text{\textbf{CFD}}\equiv1\pmod N$ and $\text{\textbf{x}}^\frac{\text{\textbf{CFD}}}{2}\neq N-1\pmod N$. If any of the denominators satisfies these three conditions, it will give us two factors.\\

%#####################
%                    \input{Conclusion}
%#####################
\section{conclusion}
In this study we have constructed a computational language for the
simulation of quantum algorithms and presented the program code for
an algorithm. We have also demonstrated that every unitary operator
has a representation in this computational language. An important
future challenge is to develop this computational language to
include the theory of  density operator.
\section{acknowledgments}
I would like to thank my supervisor Prof. Andrei Khrennikov  for
fruitful discussions on the foundations of quantum computing.  I am
also grateful to Yaroslav Volovich for his involvement and ideas in
this research.

\end{document}